# Stimulated wave of polarization in 1D Ising chain

Jae-Seung Lee and A. K. Khitrin

*Department of Chemistry, Kent State University, Kent, Ohio 44242-0001*

**Abstract**

It is demonstrated that in one-dimensional Ising chain with nearest-neighbor interactions, irradiated by a weak resonant transverse field, a stimulated wave of flipped spins can be triggered by a flip of a single spin. This analytically solvable model illustrates mechanisms of quantum amplification and quantum measurement.

## 1. INTRODUCTION

A simple logic scheme of quantum measurement [1] can be represented as

$$U|\psi_{in}\rangle = U(a|0\rangle_O + b|1\rangle_O)|0\rangle_D = a|0\rangle_O|0\rangle_D + b|1\rangle_O|1\rangle_D, \quad |a|^2 + |b|^2 = 1, \quad (1)$$

where $|0\rangle_O$ and $|1\rangle_O$ are two states of a quantum object, $|0\rangle_D$ and $|1\rangle_D$ are two macroscopically distinct states of a quantum measuring device. In the initial state $|\psi_{in}\rangle$, the object is in some arbitrary superposition state while the device is in its ground state. The unitary transformation $U$ is the controlled-NOT (CNOT) gate between two qubits representing the object and the measuring device. The CNOT gate flips the state of the second qubit when the first qubit is in the state $|1\rangle_O$ and does not do anything when the first qubit is in the state $|0\rangle_O$. More details can be added to this scheme if we suppose that the macroscopic measuring device is a composite quantum system consisting, as an example, of two-level systems 2 to N with the two macroscopically distinct states: the ground state $|0\rangle_D = |0\rangle_2|0\rangle_3\ldots|0\rangle_{N-1}|0\rangle_N$ and the state with all qubits in their excited states $|1\rangle_D = |1\rangle_2|1\rangle_3\ldots|1\rangle_{N-1}|1\rangle_N$. When qubits are implemented by spins ½, these two states are two ferromagnetic states with all spins up or all spins down. Index 1 will be used for the object: $|0\rangle_O = |0\rangle_1$, and $|1\rangle_O = |1\rangle_1$. Then, the logic scheme of quantum measurement can be written as

$$|\psi_{out}\rangle = U|\psi_{in}\rangle = U(a|0\rangle_1 + b|1\rangle_1)|0\rangle_2|0\rangle_3\ldots|0\rangle_{N-1}|0\rangle_N$$
$$= a|0\rangle_1|0\rangle_2\ldots|0\rangle_{N-1}|0\rangle_N + b|1\rangle_2|1\rangle_3\ldots|1\rangle_{N-1}|1\rangle_N. \quad (2)$$

There exist an infinite number of unitary operators which convert the state $|\psi_{in}\rangle$

into the state $|\psi_{out}\rangle$, depending on what they do to the other states of the system. As an example, in Ref. [3] a chain of CNOT operations between the first and each of the other qubits has been proposed. Another possible form is a chain of unitary CNOT gates [2]

$$U = CNOT_{N-1,N} CNOT_{N-2,N-1} \ldots CNOT_{2,3} CNOT_{1,2} \qquad (3)$$

where $CNOT_{m,n}$ negates the *n*-th qubit conditioned on the *m*-th qubit. If the first qubit is in the state $|1\rangle_1$, it flips the second qubit, then the second flips the third, and so on. This wave of flipped qubits, triggered by the first qubit, propagates until it covers the entire system. An advantage of the circuit in Eq. (3) is that it requires only interactions between neighbor qubits and, therefore, potentially can be implemented in large systems.

In practical measuring devices, the dynamics of the combined system (object + measuring device) is accompanied by decoherence resulting from interaction with an environment. In an idealized model, we can suppose that the reversible unitary evolution and the decoherence are separated in time. First, the unitary transformation of Eq. (3) takes place and then the irreversible decoherence increases the entropy and converts the pure state of Eq. (2) into the mixed state with the density matrix

$$\rho = |a|^2 |0_1 \ldots 0_N\rangle\langle 0_1 \ldots 0_N| + |b|^2 |1_1 \ldots 1_N\rangle\langle 1_1 \ldots 1_N|. \qquad (4)$$

This density matrix describes an ensemble of outcomes of individual quantum measurements. With the probability $|a|^2$ we find the measuring device in its ground state and the wavefunction of the object collapsed to the corresponding state. With the probability $|b|^2$ we find another "reading" of the measuring device and the object in its excited state. In fact, in real situations when the measuring device is a comparatively small quantum system, the two processes, unitary evolution and decoherence, can be well separated in time [2].

The most important feature of the process leading to the state of Eq. (2) is the signal amplification: polarization of a single two-level system is converted into a macroscopic total polarization of the measuring device. In classical physics, amplification is based on non-linear behavior. On the other hand, quantum mechanics is linear and the mechanisms of amplification are different [2-6]. Amplified quantum detection [4] or measurement [2,3] requires a collective dynamics, which propagates entanglements and correlations through the entire system. In the absence of long-range interactions, the polarization wave, represented by the unitary transform of Eq. (3), is the fastest and the most efficient way of creating entanglements. As an example, if the initial state of the object is the superposition $2^{-1/2}(|0\rangle_O + |1\rangle_O)$, the chain of gates in Eq. (3) creates the maximally entangled "Schrödinger cat" state of the entire system.

Implementation of each of the CNOT gates in Eq. (3) requires coherent control and addressing of individual qubits. On the other hand, there exists a Hamiltonian $H = (i/\tau)\ln U$ which gives the unitary evolution of Eq. (3) after the evolution time $\tau$, although the structure of this Hamiltonian might be very complex. In this work, we propose an analytically solvable model with a simple Hamiltonian, which exhibits quantum dynamics similar to the ideal scheme of Eq. (3) and, therefore, realizes a mechanism of quantum amplification.

## 2. MODEL

Let us consider a one-dimensional Ising chain with nearest-neighbor interactions, irradiated at resonance by a weak transverse monochromatic field. The Hamiltonian of the system is

$$H = \frac{\omega_0}{2}\sum_{i=1}^{N}\sigma_i^z + \omega_1\sum_{i=1}^{N}\sigma_i^x \cos\omega_0 t + \frac{J}{4}\sum_{i=1}^{N-1}\sigma_i^z\sigma_{i+1}^z, \qquad (5)$$

where $\omega_0$ is the energy difference ($\hbar = 1$) between the excited and ground states of an isolated spin (qubit), $J$ is the interaction constant, $\omega_1 \ll J \ll \omega_0$ is the amplitude of irradiation, $\sigma^z$ and $\sigma^x$ are the Pauli operators. The principle of the operation of this model is illustrated in Fig.1. When a spin is at the either end of the chain or when it has two neighbors in the same state, interaction with the neighbor(s) makes the spin off-resonant and the irradiation field does not change its state. When the two neighbors are in different states, the resonant irradiation field flips the spin. Therefore, if all spins are in the same state, the state of the entire system is stationary. If the first spin is flipped, its neighbor becomes resonant and flips, then the next spin flips, and so on. Quantum-mechanical solution for this dynamics is given in the next sections.

## 3. SECULAR HAMILTONIAN

In the rotating frame at $\omega_1 \ll J \ll \omega_0$, one can neglect the non-resonant counter-rotating component of the transverse field. Then, the Hamiltonian in the rotating frame becomes

$$H_{\text{rot}} = H_x + H_{zz} = \frac{\omega_1}{2}\sum_{i=1}^{N}\sigma_i^x + \frac{J}{4}\sum_{i=1}^{N-1}\sigma_i^z\sigma_{i+1}^z. \qquad (6)$$

It is the Hamiltonian of an Ising chain in a transverse field [7]. At $\omega_1 \ll J$, we can restrict ourselves by considering only the secular part of the Hamiltonian of Eq. (6), which is the time-independent part of the Hamiltonian in the interaction representation: $\widetilde{H}(t) = \exp(-iH_{zz}t)H_x \exp(iH_{zz}t)$. By noticing that the terms in $H_{zz}$ commute with each other and that $\exp\left(-i\frac{J}{4}\sigma_i^z\sigma_{i+1}^z t\right) = I\cos\left(\frac{Jt}{4}\right) - i\sigma_i^z\sigma_{i+1}^z \sin\left(\frac{Jt}{4}\right)$, where $I$ is the identity operator, one obtains

$$\begin{aligned}
\widetilde{H}(t) &= \exp(-iH_{zz}t)H_x \exp(iH_{zz}t) \\
&= \exp\left[-i(Jt/4)\sum_{i=1}^{N-1}\sigma_i^z\sigma_{i+1}^z\right]\frac{\omega_1}{2}\sum_{i=1}^{N}\sigma_i^x \exp\left[i(Jt/4)\sum_{i=1}^{N-1}\sigma_i^z\sigma_{i+1}^z\right] \\
&= \frac{\omega_1}{2}\sum_{i=2}^{N-1}\exp[-i(Jt/4)\sigma_{i-1}^z\sigma_i^z]\exp[-i(Jt/4)\sigma_i^z\sigma_{i+1}^z]\sigma_i^x \\
&\quad \times \exp[i(Jt/4)\sigma_i^z\sigma_{i+1}^z]\exp[i(Jt/4)\sigma_{i-1}^z\sigma_i^z] \\
&\quad + \frac{\omega_1}{2}\exp[-i(Jt/4)\sigma_1^z\sigma_2^z]\sigma_1^x \exp[i(Jt/4)\sigma_1^z\sigma_2^z] \\
&\quad + \frac{\omega_1}{2}\exp[-i(Jt/4)\sigma_{N-1}^z\sigma_N^z]\sigma_N^x \exp[i(Jt/4)\sigma_{N-1}^z\sigma_N^z] \\
&= \frac{\omega_1}{4}\sum_{i=2}^{N-1}\left[(\sigma_i^x - \sigma_{i-1}^z\sigma_i^x\sigma_{i+1}^z) + (\sigma_i^x + \sigma_{i-1}^z\sigma_i^x\sigma_{i+1}^z)\cos Jt + (\sigma_{i-1}^z\sigma_i^y + \sigma_i^y\sigma_{i+1}^z)\sin Jt\right] \\
&\quad + \frac{\omega_1}{2}\left[\sigma_1^x \cos\frac{Jt}{2} + \sigma_1^y\sigma_2^z \sin\frac{Jt}{2} + \sigma_N^x \cos\frac{Jt}{2} + \sigma_{N-1}^z\sigma_N^y \sin\frac{Jt}{2}\right],
\end{aligned} \quad (7)$$

The time-independent part of this expression is the secular Hamiltonian

$$\widetilde{H}_{\text{secular}} = \frac{\omega_1}{4}\sum_{i=2}^{N-1}\sigma_i^x\left(1 - \sigma_{i-1}^z\sigma_{i+1}^z\right). \quad (8)$$

This Hamiltonian has exactly the same form as the model Hamiltonian used to describe a motion of domain walls in 1D Ising magnet [8]. The terms of the three-spin effective Hamiltonian of Eq. (8) have a straightforward interpretation: the spin-flipping operator $\sigma_i^x$ for the spin $i$ is "turned off" when its two neighbors are in the same state and $\left(1 - \langle\sigma_{i-1}^z\sigma_{i+1}^z\rangle\right)$ is zero.

## 4. EIGENSPACE

There are several successful examples, when simplification of spin [9] or spatial [10] parts of a Hamiltonian has led to analytically solvable models with non-trivial spin dynamics. Some dynamical problems in one-dimensional chains can be solved exactly [11]. For the model we consider in this work, simplification arises from a special type of initial conditions.

We are interested in the dynamics, which starts with one of the two initial states: when all the qubits are in the ground state and when the first qubit is flipped. In this case, evolution is confined within a small subspace of the entire Hilbert space. Dimensionality of this subspace is $N+1$, compared to $2^N$ of the entire space.

Let $|\Psi_k\rangle$ be the state with the first $k$ spins of the chain flipped (i.e. in the state down). By repeatedly acting with the secular Hamiltonian of Eq. (8) on $|\Psi_k\rangle$, one can see that there exist three subspaces spanned by $|\Psi_k\rangle$'s.

$$H_{\text{secular}}|\Psi_k\rangle = \begin{cases} 0 & \text{if} \quad k=0, N \\ \frac{\omega_1}{2}|\Psi_2\rangle & \text{if} \quad k=1 \\ \frac{\omega_1}{2}(|\Psi_{k-1}\rangle+|\Psi_{k+1}\rangle) & \text{if} \quad 2 \le k \le N-2 \\ \frac{\omega_1}{2}|\Psi_{N-2}\rangle & \text{if} \quad k=N-1 \end{cases} \qquad (9)$$

The dynamics is very different for the two slightly different initial states $|\Psi_0\rangle$ and $|\Psi_1\rangle$. The state $|\Psi_0\rangle$ does not change since it is an eigenstate of the secular

Hamiltonian, while the state $|\Psi_1\rangle$ can be converted into the states with multiple flipped qubits (up to $N$-1). There are no non-zero matrix elements of the Hamiltonian of Eq. (8) between $|\Psi_k\rangle$'s and any other vectors of the multiplicative basis. From Eq. (9), one concludes that the secular Hamiltonian of Eq. (8), in the subspace spanned by $|\Psi_k\rangle$'s with $k = 1, \ldots N - 1$, has non-zero elements only on the first super- and sub-diagonals.

## 5. SOLUTION AND RESULTS

Let $M_N$ be a $N \times N$ tridiagonal matrix with $(M_N)_{i,i} = a$, $(M_N)_{i,i+1} = (M_N)_{i+1,i} = b$, where $a$ and $b$ are real numbers [12]. The eigenvalues can be obtained by solving the equation $\det(M_N - \lambda I_N) = 0$, where $\lambda$ and $I_N$ are the eigenvalue and the $N \times N$ identity matrix, respectively. It is easy to check that the calculation of the determinant produces a recursion relation

$$\det(M_N - \lambda I_N) = (a - \lambda)\det(M_{N-1} - \lambda I_{N-1}) - b^2 \det(M_{N-2} - \lambda I_{N-2}),$$

which can be solved to give $\det(M_N - \lambda I_N) = (-1)^N \gamma_-^N \frac{1 - (\gamma_+/\gamma_-)^{N+1}}{1 - (\gamma_+/\gamma_-)}$, where $\gamma_\pm = \frac{1}{2}\left(\lambda - a \pm \sqrt{(\lambda - a)^2 - 4b^2}\right)$. The nontrivial solution satisfying $\det(M_N - \lambda I_N) = 0$ can be found from $(\gamma_+/\gamma_-)^{N+1} = 1$, which gives the eigenvalues $\lambda_p = a - 2b\cos\frac{p\pi}{N+1}$ with $p = 1, \cdots, N$. Let $\vec{\alpha}_p = (\alpha_{p1}, \cdots, \alpha_{pN})$ be the eigenvector corresponding to $\lambda_p$. From $M_N \vec{\alpha}_p = \lambda_p \vec{\alpha}_p$, $\alpha_{p,k+1} + \alpha_{p,k-1} = -2\alpha_{p,k}\cos\frac{p\pi}{N+1}$, $\alpha_{p,2} = -2\alpha_{p,1}\cos\frac{p\pi}{N+1}$, and $\alpha_{p,N-1} = -2\alpha_{p,N}\cos\frac{p\pi}{N+1}$. The solution of these equations gives $\alpha_{p,k} = (-1)^{k-1}\frac{\sin[pk\pi/(N+1)]}{\sin[p\pi/(N+1)]}$. From the normalization condition $\sum_{k=1}^{N}|\alpha_{p,k}|^2 = 1$ and

$$\sum_{k=1}^{N}\sin^{2}\left(\frac{pk\pi}{N+1}\right)=\frac{1}{2}(N+1), \quad \text{we obtain} \quad \alpha_{p,k}=(-1)^{k-1}\sqrt{\frac{2}{N+1}}\sin\left(\frac{pk\pi}{N+1}\right), \quad \text{where}$$

$p=1,\cdots,N$, and $k=1,\cdots,N-1$.

From the above results, the secular Hamiltonian of Eq. (8), in the subspace spanned by $|\Psi_k\rangle$'s, has the eigenvalues

$$\lambda_p = -\omega_1 \cos\frac{p\pi}{N} \tag{10}$$

and corresponding eigenvectors

$$|\Phi_p\rangle = \sqrt{\frac{2}{N}}\sum_{k=1}^{N-1}(-1)^{k-1}\sin\frac{pk\pi}{N}|\Psi_k\rangle, \tag{11}$$

where $p=1,\cdots,N-1$.

While $|\Psi_0\rangle$ is stationary under the secular Hamiltonian of Eq. (8), $|\Psi_1\rangle$ initiates a very interesting "quantum domino" dynamics which results in macroscopic changes. For the operator of the total polarization $P=\sum_{k=1}^{N}\sigma_k^z$, $\langle\Psi_l|P|\Psi_k\rangle=(N-2k)\delta_{kl}$. The time dependence of its average value is given by

$$\begin{aligned}\langle P(t)\rangle &= \text{tr}\left\{Pe^{-i\widetilde{H}_{\text{secular}}t}|\Psi_1\rangle\langle\Psi_1|e^{i\widetilde{H}_{\text{secular}}t}\right\} \\ &= \sum_{k,l,p,q,r,s=1}^{N-1}\langle\Psi_k|\Phi_p\rangle\langle\Phi_p|e^{-i\widetilde{H}_{\text{secular}}t}|\Phi_q\rangle\langle\Phi_q|\Psi_1\rangle\langle\Psi_1|\Phi_r\rangle\langle\Phi_r|e^{i\widetilde{H}_{\text{secular}}t}|\Phi_s\rangle \\ &\quad \times \langle\Phi_s|\Psi_l\rangle\langle\Psi_l|P|\Psi_l\rangle.\end{aligned} \tag{12}$$

Using Eqs. (10) and (11), Eq. (12) can be written explicitly as

$$\langle P(t)\rangle = \frac{4}{N^2}\sum_{k=1}^{N-1}\sum_{p,r=1}^{N-1}(N-2k)\sin\frac{pk\pi}{N}\sin\frac{p\pi}{N}\sin\frac{rk\pi}{N}\sin\frac{r\pi}{N}\cos\left[\left(\cos\frac{p\pi}{N}-\cos\frac{k\pi}{N}\right)\omega_1 t\right]. \tag{13}$$

We can also calculate polarizations of individual spins

$$p_m = \langle\Psi_l|\sigma_m^z|\Psi_k\rangle = \begin{cases}\delta_{kl} & \text{if } k<m \\ -\delta_{kl} & \text{if } k\geq m\end{cases}.$$

The time dependence of its average value is given by

$$p_m(t) = \text{tr}\{\sigma_m^z \exp(-i\tilde{H}_{sec ular}t)|\Psi_1\rangle\langle\Psi_1|\exp(i\tilde{H}_{sec ular}t)\}$$

$$= \frac{4}{N^2}\left(\sum_{k<m} - \sum_{k\geq m}\right)\sum_{p,r=1}^{N-1} \sin\frac{pk\pi}{N}\sin\frac{p\pi}{N}\sin\frac{rk\pi}{N}\sin\frac{r\pi}{N}\cos\left[\left(\cos\frac{p\pi}{N} - \cos\frac{k\pi}{N}\right)\omega_1 t\right].$$

(14)

In order to independently check the results summarized by Eqs. (13) and (14), we performed a computer simulation of a small eight-spin chain with the secular Hamiltonian (Eq. (8)). The simulation was done in the entire Hilbert space of the eight-spin system. The simulated dynamics of polarizations coincided precisely with those given by the analytical expressions of Eqs. (13) and (14).

The "snapshots" of polarizations $p_m = \langle\sigma_m^z(t)\rangle$ at $\omega_1 t$ = 0, 25, 50, 75, 100, and 105.7 for $N$ = 100 are displayed in Fig. 2. One can clearly see the wave of flipped qubits, propagating from the left end of the chain. The width of the transition region, or the "domain wall", also increases with time.

Amazingly, this dynamics gives practically linear time dependence of the total polarization, till the wave reaches the opposite end of the chain, where the wave reflects and moves back. The changes in the total polarization $\langle\Delta P(t)\rangle = \langle P(t)\rangle - \langle P(0)\rangle$ are plotted in Fig. 3 for $N$ = 25, 50, 75, and 100 as functions of the dimensionless time $\omega_1 t$.

## 6. DISCUSSION

For large $N$, the maximum change of polarization is reached at $\omega_1 t \approx 1.06\ N$. At this time, the absolute value of the total polarization is about 87% of its maximum value. By comparing the change of the polarization produced by the flip of a single qubit and the maximum change of the total polarization, resulting from the dynamics triggered by a flip of one qubit, one can introduce a coefficient of amplification $\alpha = |\Delta P|/2$. For $N \gg 1$, in our ideal model the coefficient of amplification $\alpha \approx 0.87 N$ can be arbitrary large for long chains. Another related characteristic is the contrast $C = \Delta P / P(0)$, introduced in Ref. [3], which is a relative change of polarization (magnetization) of the entire spin system. Its maximum possible value is $C = 2$. In our model, for $N \gg 1$, $C \approx 1.73$.

Propagation of a stimulated polarization wave, studied in this work, is a reversible quantum dynamics. The evolution can be reversed at any time by changing the sign of the effective Hamiltonian of Eq. (8). Since the effective Hamiltonian is linear in $\omega_1$, the change of its sign can be easily accomplished by a 180° shift of the phase of the resonant irradiation.

A similar quantum dynamics of amplification can be used for designing efficient detectors or quantum measuring devices. However, we do not expect that analytical solutions will be available for systems with more realistic Hamiltonians, and computer simulations will, probably, be restricted to comparatively small composite quantum systems. In addition, relaxation, decoherence, and distribution of the initial states would affect operation of real devices. The role of these factors in the amplification dynamics has not been studied yet.

**ACKNOWLEDGEMENTS**

This work was supported by Kent State University and US - Israel Binational Science Foundation.

**REFERENCES**

[1] J. P. Paz and W. H. Zurek, Environment-induced decoherence and the transition from quantum to classical, in *Fundamentals of quantum information : quantum computation, communication, decoherence and all that,* edited by D. Heiss (Springer, Berlin, 2002) pp. 77-148.

[2] J.-S. Lee and A. K. Khitrin, *submitted to Phys. Rev. Letters.*

[3] P. Cappellaro, J. Emerson, N. Boulant, C. Ramanathan, S. Lloyd, and D. G. Cory, arXiv:quant-ph/0411128 (2004).

[4] J.-S. Lee and A. K. Khitrin, J. Chem. Phys. **121**, 3949 (2004).

[5] M. Kindermann and D. G. Cory, arXiv:quant-ph/0411038 (2004).

[6] D. P. DiVincenzo, J. Appl. Phys. **85**, 4785 (1999).

[7] B. K. Chakrabarti, A. Dutta, and P. Sen, *Quantum Ising Phases and Transitions in Transverse Ising Models* (Springer-Verlag, Berlin, 1996).

[8] V. Subrahmanyam, Phys. Rev. B **68**, 212407 (2003).

[9] A. K. Khitrin, Phys. Lett. A, **214**, 81 (1996).

[10] M. G. Rudavets and E. B. Fel'dman, JETP Letters, **75**, 635 (2002).

[11] S. I. Doronin, I. I. Maksimov, and E. B. Fel'dman, JETP, **91**, 597 (2000); E. B. Fel'dman and M. G. Rudavets, arXiv:cond-mat/0411613 (2004).
12

[12] The case $N \to \infty$ has been studied in the context of the continuous-time quantum walk: A. M. Childs, R. Cleve, E. Deotto, E. Farhi, S. Gutmann, and D. A. Spielman, in *Proceedings of the 35th ACM Symposium on Theory of Computing* (ACM, New York, 2003), p. 59.

**Figure captions**

Fig. 1. 1D Ising chain under resonant irradiation.

Fig. 2. Snapshots of individual cites' polarizations at $\omega_1 t = 0$ (a), 25 (b), 50 (c), 75 (d), 100 (e), and 105.7 (f) for $N = 100$.

Fig. 3. Change of the total polarization $\Delta P(t)$ for $N = 25$ (dash-dot line), 50 (dotted line), 75 (dashed line), and 100 (solid line).

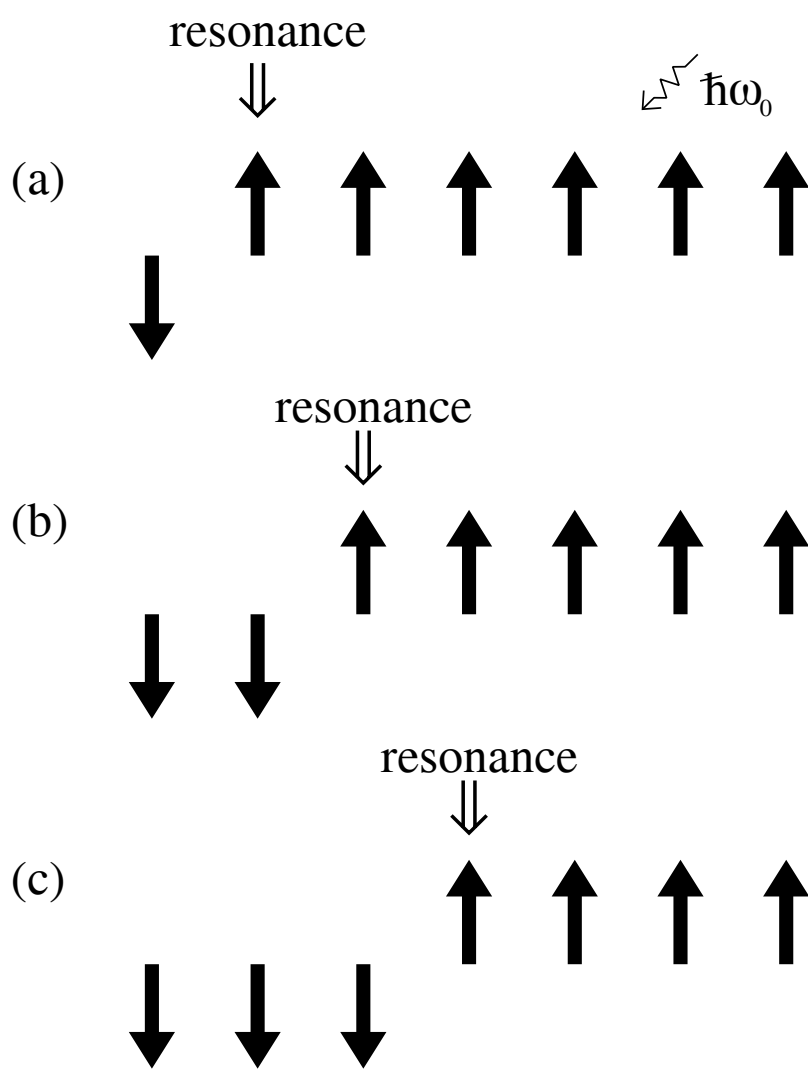

Fig. 1

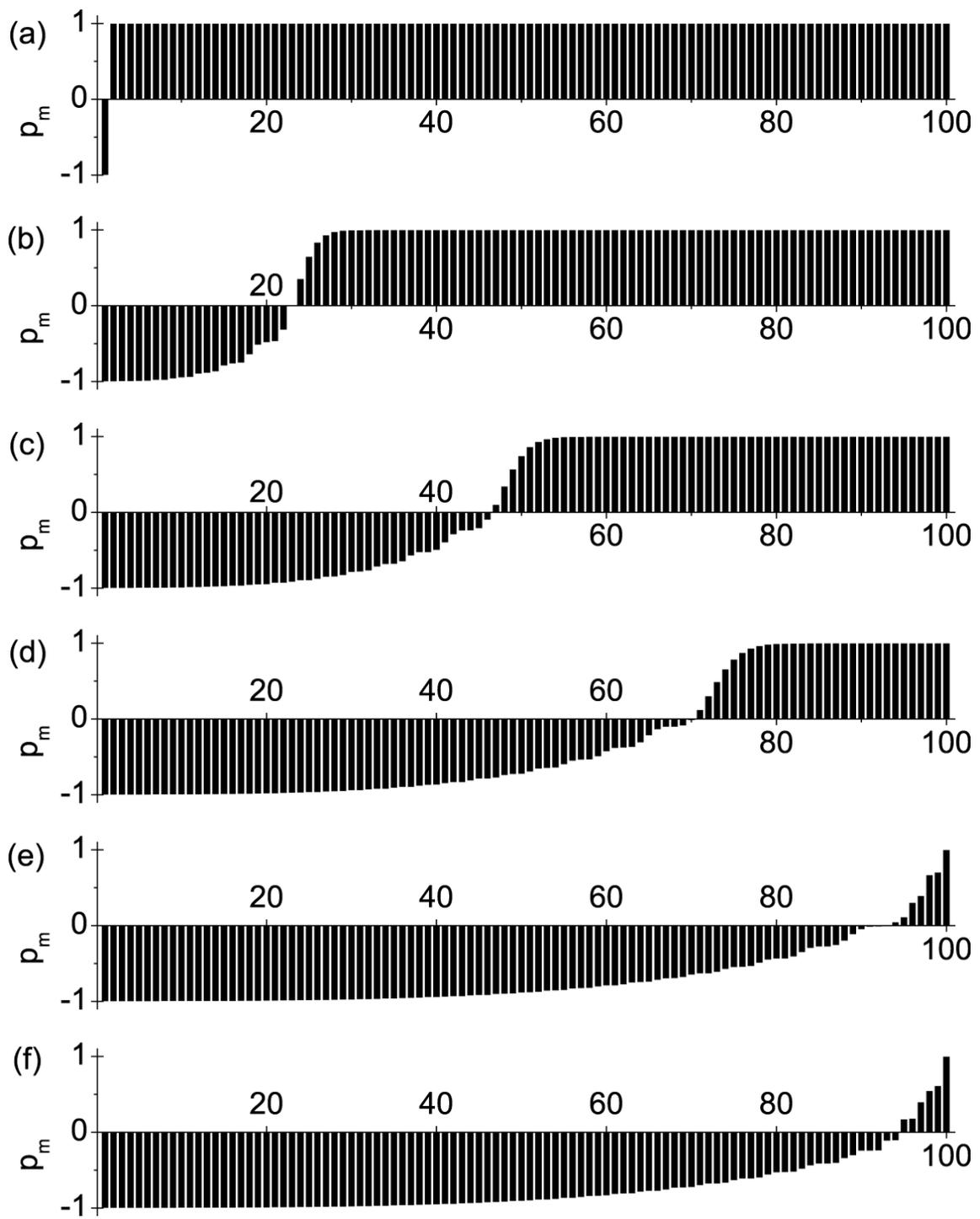

Fig. 2

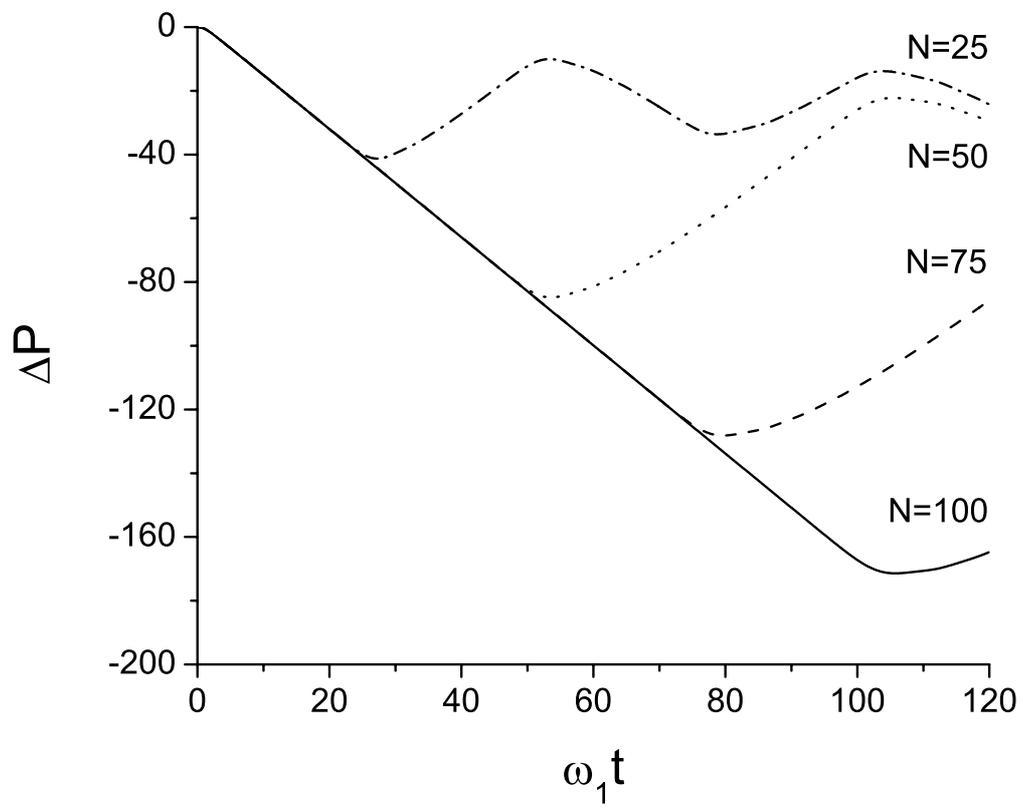

Fig. 3